\documentclass[twocolumn]{aastex6}
\usepackage{epsfig}
\usepackage{natbib}
\usepackage{amsmath}
\usepackage{color}
\usepackage{graphicx}

\newcommand{\kms}{km~s$^{-1}$}

\DeclareMathAlphabet{\mathpzc}{OT1}{pzc}{m}{it}

\begin{document}
\title{The Distance to the Galaxy Coma P}

\author{Gagandeep S. Anand}
\and
\author{R. Brent Tully}
\affil{Institute for Astronomy, University of Hawaii, 2680 Woodlawn Drive, Honolulu, HI 96822, USA}
\and
\author{Igor D. Karachentsev}
\and
\author{Dmitry I. Makarov}
\and
\author{Lidia Makarova}
\affil{Special Astrophysical Observatory, Nizhniy Arkhyz, Karachai-Cherkessia 369167, Russia}
\and
\author{Luca Rizzi}
\affil{W. M. Keck Observatory, 65-1120 Mamalahoa Hwy., Kamuela, HI 96743, USA}
\and
\author{Edward J. Shaya}
\affil{Astronomy Department, University of Maryland, College Park, MD 20743, USA}

\begin{abstract}
If the extremely low surface brightness galaxy Coma~P lies at $5.5\pm0.3$~Mpc as recently proposed then it would have an extraordinarily deviant peculiar velocity of $\sim 900$~\kms\ at a location where differential velocities between galaxies are low.  We have accessed the images from the HST archives used to derive the literature distance from the magnitude of the tip of the red giant branch.  Our analysis gives the distance to be $10.9\pm1.0$~Mpc.  At this location the galaxy lies within the infall region of the Virgo Cluster, such that its still considerable peculiar velocity  of $\sim 500$~\kms\ is consistent with an established model.  Coma~P has an unusually pronounced asymptotic giant branch relative to its red giant branch.  The dominant stellar population is just a few Gyr old.

\end{abstract}

\smallskip
\section{Introduction}

Recently, attention has been given to the ``almost dark" low surface brightness galaxy Coma~P (AGC 229385 = PGC 5809449).  The object was detected as an HI source in the blind survey of the extragalactic sky ALFALFA  \citep{2011AJ....142..170H} with no counterpart in the Sloan Digital Sky Survey but the suggestion of a detection in the GALEX archival images.  Subsequently, \citet{2015ApJ...801...96J} identified an extremely low surface brightness galaxy at the location of the HI source with the WIYN 3.5m telescope and resolved the galaxy at HI with observations using the Westerbork Synthesis Radio Telescope.  \citet{2018AJ....155...65B} obtained kinematic and imaging information at higher resolution with the Very Large Array.  \citet{2017AAS...23021403B} report on optical imaging with Hubble Space Telescope (HST) and give a distance of $5.5\pm0.3$~Mpc from resolved stars.  These various authors discuss the interesting properties of Coma~P, some of which depend on the assumed distance. Here we suggest that asymptotic giant branch (AGB) stars have been mistakenly identified as red giant branch (RGB) stars with the result that the distance has been underestimated by roughly a factor of two.  The revised distance has important implications for the peculiar velocity of Coma~P and intriguing implications regarding the ages of stars in this poorly understood class of galaxies.

Coma~P has an extremely precise heliocentric velocity of $1348\pm1$~\kms\ from the HI observations, which translates to 1296~\kms in the Local Sheet frame (the \citet{2008ApJ...676..184T} variation on the Local Group frame) and 1320~\kms\ in the frame of our galaxy \citep{2012ApJ...753....8V}.  All other galaxies in the projected vicinity of Coma~P at roughly the Brunker et al. distance of 5.5 Mpc have velocities $\sim 400$~\kms\ in accordance with cosmic expansion, with radial dispersions of only 85~\kms\ \citep{2003A&A...398..479K}.  The implied peculiar velocity if the distance is 5.5~Mpc is $\sim +884$~\kms.  This extraordinary situation caught our attention and caused us to re-evaluate the evidence for the distance to Coma~P.

\section{The TRGB Distance}

The tip of the red giant branch (TRGB) methodology \citep{1993ApJ...417..553L} based on imaging with HST can regularly give distances with a reliability of 5\% for targets within 10~Mpc \citep{2009ApJ...690..389M, 2009AJ....138..332J, 2014ApJ...785....3M, 2017ApJ...835...28J}.  In rare cases when the procedure fails badly it is inevitably because the asymptotic giant branch (AGB) has been mistaken for the red giant branch (RGB) \citep{2007ApJ...667L.151A}.

\begin{figure*}
\plotone{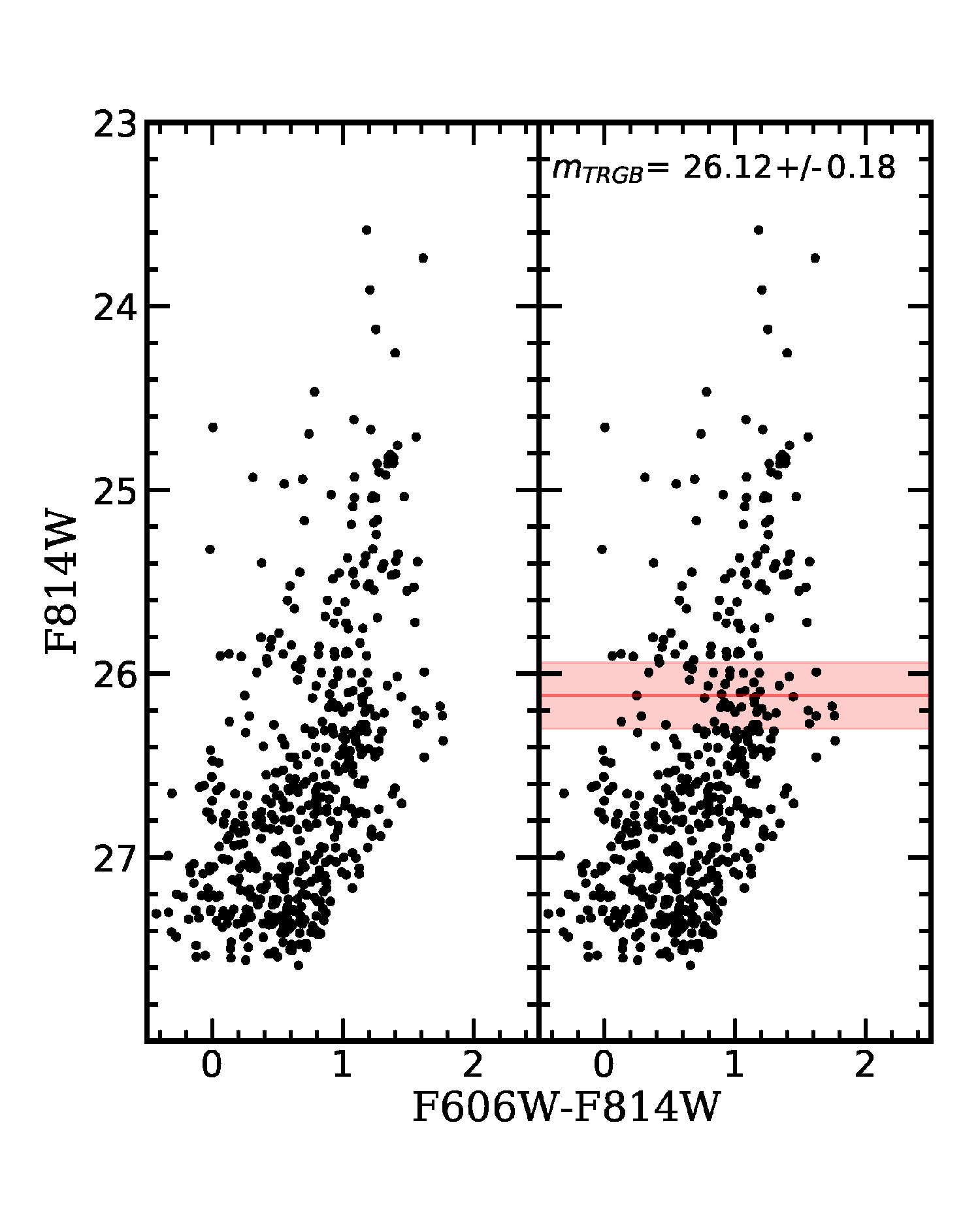}
\caption{Color magnitude diagram of stars in close proximity of Coma P.  {\it Left:} uninterpreted. {\it Right:} TRGB fit with $1\sigma$ uncertainty.}
\label{cmd}
\end{figure*}

To investigate the matter, we acquired the HST images for program GO-14108 (PI: J. Salzer) from the Mikulski Archive for Space Telescopes. Observations were made over two orbits with roughly equal $\sim 2600$~s integrations in the F606W and F814W bands using the Advanced Camera for Surveys (ACS). We carried out our standard analysis: stellar photometry using DOLPHOT   \citep{2000PASP..112.1383D, 2016ascl.soft08013D}, a maximum likelihood determination of the TRGB including recovery of synthetic stars to monitor completion and photometric uncertainties \citep{2006AJ....132.2729M}, and zero point calibration \citep{2007ApJ...661..815R}. Our color-magnitude diagram as reduced by GA and LR is seen without commentary in the left panel of Figure~\ref{cmd}.

The basis for the fit of the TRGB by \citet{2017AAS...23021403B} at F814W$^{trgb} = 24.64$ is apparent.  On closer inspection, though, it is seen that there is a step to increased density of red stars at $\sim 26$.  We posit that the true TRGB lies near this magnitude and that the red stars brighter than this level are associated with the AGB.  
We carried out two independent analyses to determine the value of the TRGB, using similar procedures on the same archival data but at two sites.  The two determinations are F814W$^{trgb} = 26.18\pm0.17$ (GA and LR) and F814W$^{trgb} = 26.06\pm0.18$ (LM and DM).  The two measures are consistent and we accept the average of F814W$^{trgb} = 26.12\pm0.18$.
Our best fit is illustrated in the right panel of Figure~\ref{cmd}.   The corresponding distance for the galaxy is $10.9\pm1.0$~Mpc. Photometry files and color magnitude diagram fits are available at the Extragalactic Distance Database\footnote{http://edd.ifa.hawaii.edu} (PGC = 5809449). Our assumption resolves several outstanding problems but, if correct, implies a numeric ratio of AGB stars to RGB stars that is unfamiliar.  

\section{Implications for the Peculiar Velocity of Coma P}

\begin{figure*}[]
\includegraphics{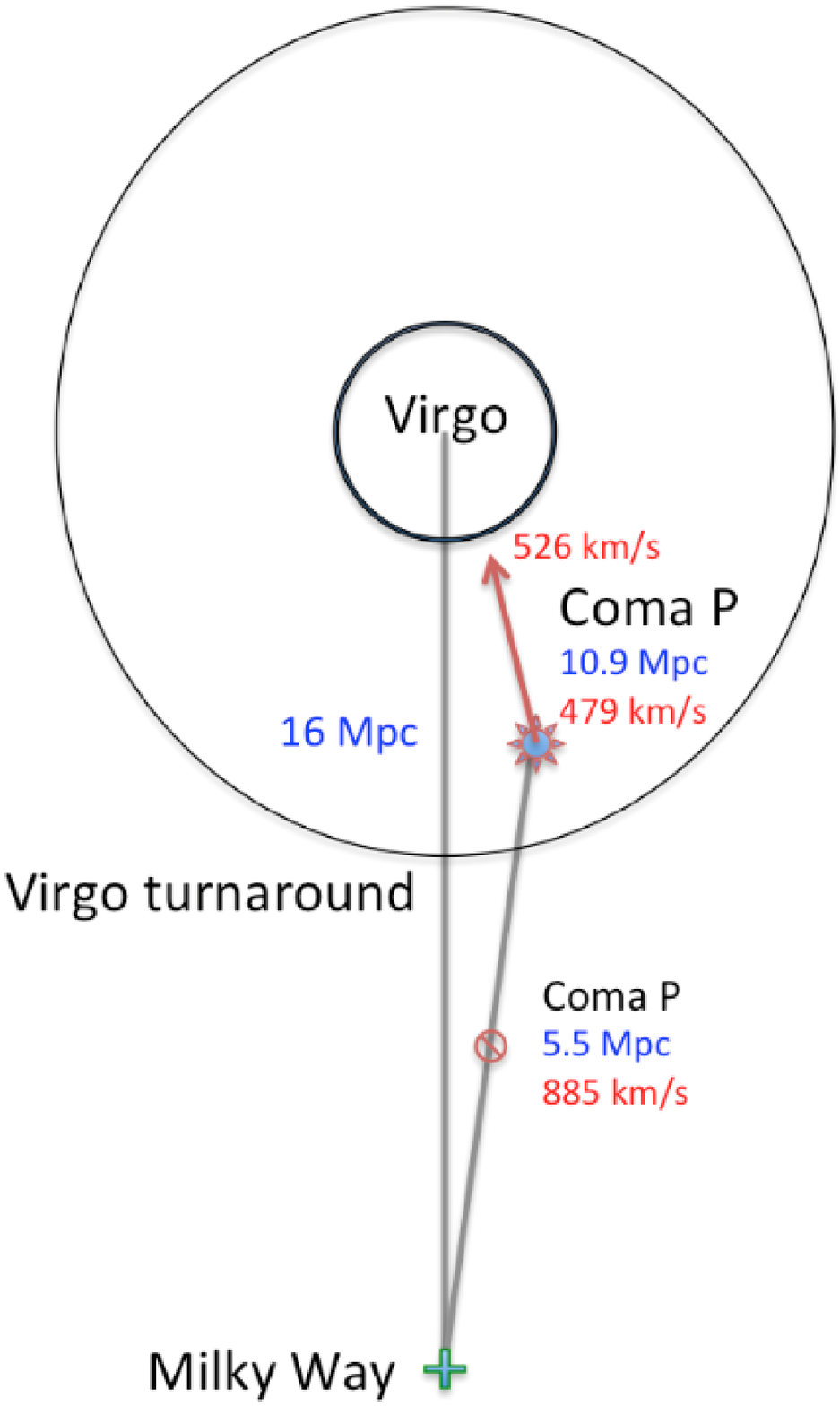}
\includegraphics{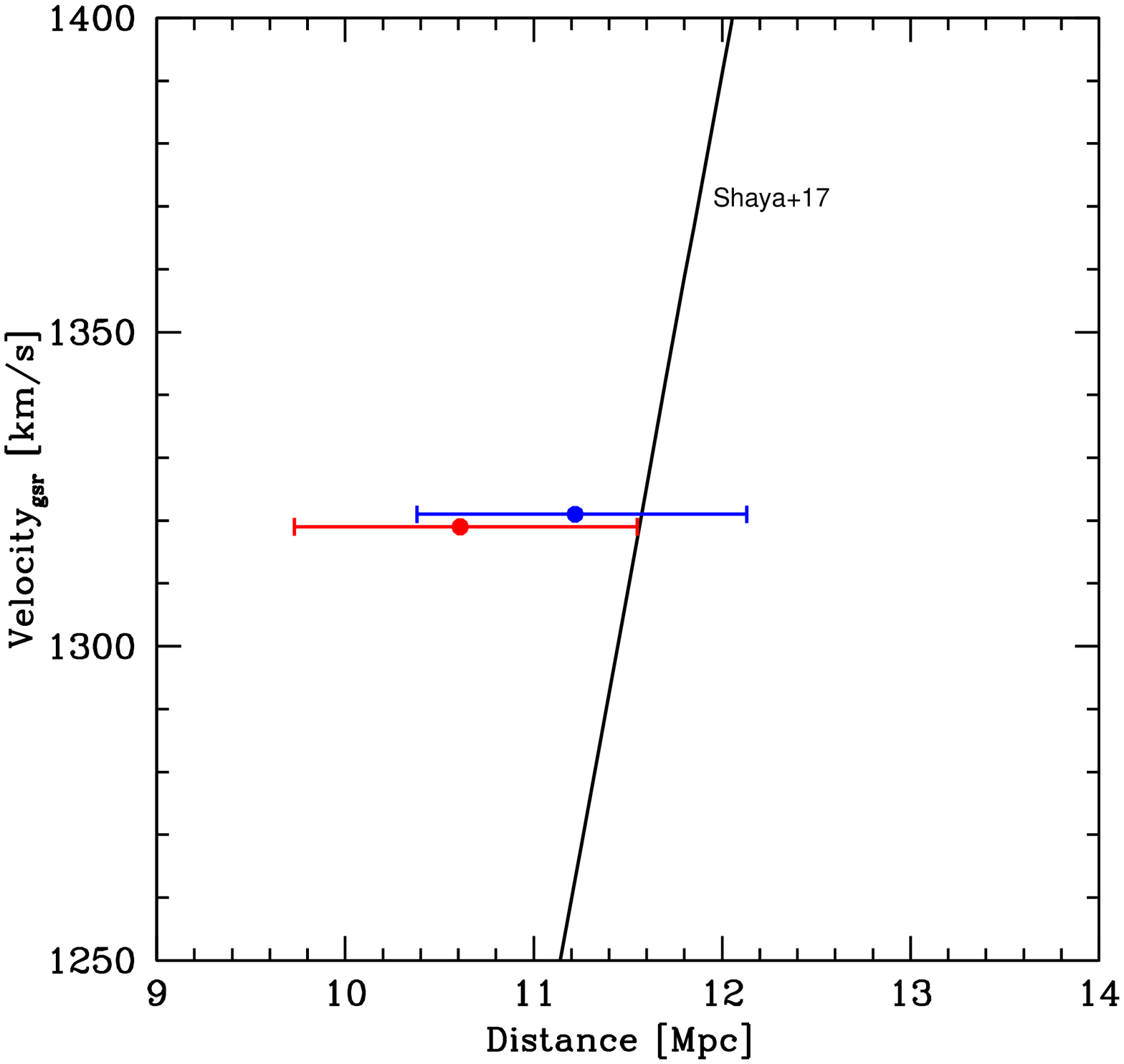}
\vspace{60mm}
\caption{{\it Left.} Schematic of the location of Coma~P in relation to the Virgo Cluster.  The heavy black circle outlines the radius of second turnaround (roughly the radius of the virialized regime) of the cluster, while the lighter ellipse defines the limits of the domain of infall into the cluster.  This domain extends to within 9 Mpc of the Milky Way.  Coma~P lies at the position of the 8-point star, 5.4 Mpc from the center of the Virgo Cluster.  The observed peculiar velocity with respect to Hubble flow is 479~\kms\ and the inferred velocity of radial infall into the cluster is 526~\kms.  If the distance of Coma~P is 5.5~Mpc it would lie at the nearer position to the Milky Way along the same line of sight and have a peculiar velocity of $\sim885$~\kms. {\it Right.} The two determinations of the distance of Coma~P, separately by GA/LR (blue) and LM/DM (red).  The observed velocity of Coma~P in the Galactic standard of rest is 1320~\kms.  The solid line is the track of velocities in this reference frame with distance from the numerical action orbit reconstruction model of \citet{2017ApJ...850..207S}.}
\label{geometry}
\end{figure*}

As mentioned in the introduction, if Coma~P lies at 5.5~Mpc then its systemic velocity would be a great mystery.  By contrast, with a distance of 10.9~Mpc its velocity is easily understood.  The situation is summarized in Figure~\ref{geometry}.  Coma~P lies at an angle of $8.0^{\circ}$ from the center of the Virgo Cluster or $2.1^{\circ}$ beyond the radius of the second turnaround radius that approximates the virial domain of the cluster \citep{2017ApJ...843...16K}. The details of the velocity field between our position and the Virgo Cluster have recently received extensive attention \citep{2014ApJ...782....4K, 2018arXiv180400469K, 2017ApJ...850..207S}.  Galaxies in an extended region around the Virgo Cluster have decoupled from cosmic expansion and are falling toward the cluster, inevitably to reach the cluster within a Hubble time.  The zero velocity surface that separates infall from expansion is mapped sufficiently to know it is not round: it is slightly squished on the axis at $90^{\circ}$ to our line of sight to Virgo in the plane of the figure and extended on the axis at $90^{\circ}$ out of the page relative to the line of sight axis \citep{2017ApJ...850..207S}.  The exact shape of the zero velocity surface is a detail but rather assuredly Coma~P is within this zone. 

In the right panel of Figure~\ref{geometry} there is a comparison of the velocity and our two alternative measures of the distance of Coma~P with expectation values from the numerical action orbit model of \citet{2017ApJ...850..207S}.  Velocities are reported in the rest frame of the center of our galaxy \citep{2012ApJ...753....8V} as necessitated by the numerical action analysis.  The velocities of the two measures are offset in the figure for clarity.

Our distance of 10.9~Mpc to Coma~P, the average of the two determinations, is consistent with the numerical action model of Shaya et al., especially given that local dispersion in velocities is expected about the mean relation.  As an aside, since Coma~P is within the Virgo infall domain there are three locations consistent with the observed velocity \citep{1981ApJ...246..680T}: a second at roughly the cluster distance and a third beyond the infall zone on the far side.  The properties of the color-magnitude diagram seen in Figure~\ref{cmd} exclude these latter two possibilities.

Two known galaxies with similar velocities lie in close proximity to Coma~P (and a third is further removed).  NGC~4561 ($V_{hello}=1396$~\kms) and IC~3605 ($V_{hello}=1362$~\kms) are considered by \citet{2017ApJ...843...16K} to be bound in halo 42020 (the Principal Galaxies Catalog identification of NGC~4561) at $1.4^{\circ}$ from Coma~P (270 kpc in projection at our Coma~P distance).  Lacking a reliable distance to this pair, \citet{2017ApJ...843...16K} placed them at a location in the Hubble flow (17.7~Mpc).  We predict that these galaxies all have similar distances, such that they lie on the near side of Virgo within the infall domain.  The slightly more removed galaxy is AGC  742507 with $V_{hello}=1209$~\kms\ and is probably in a similar situation.

\section{Implications for the Stellar Population of Coma P}

\begin{figure*}[]
\plotone{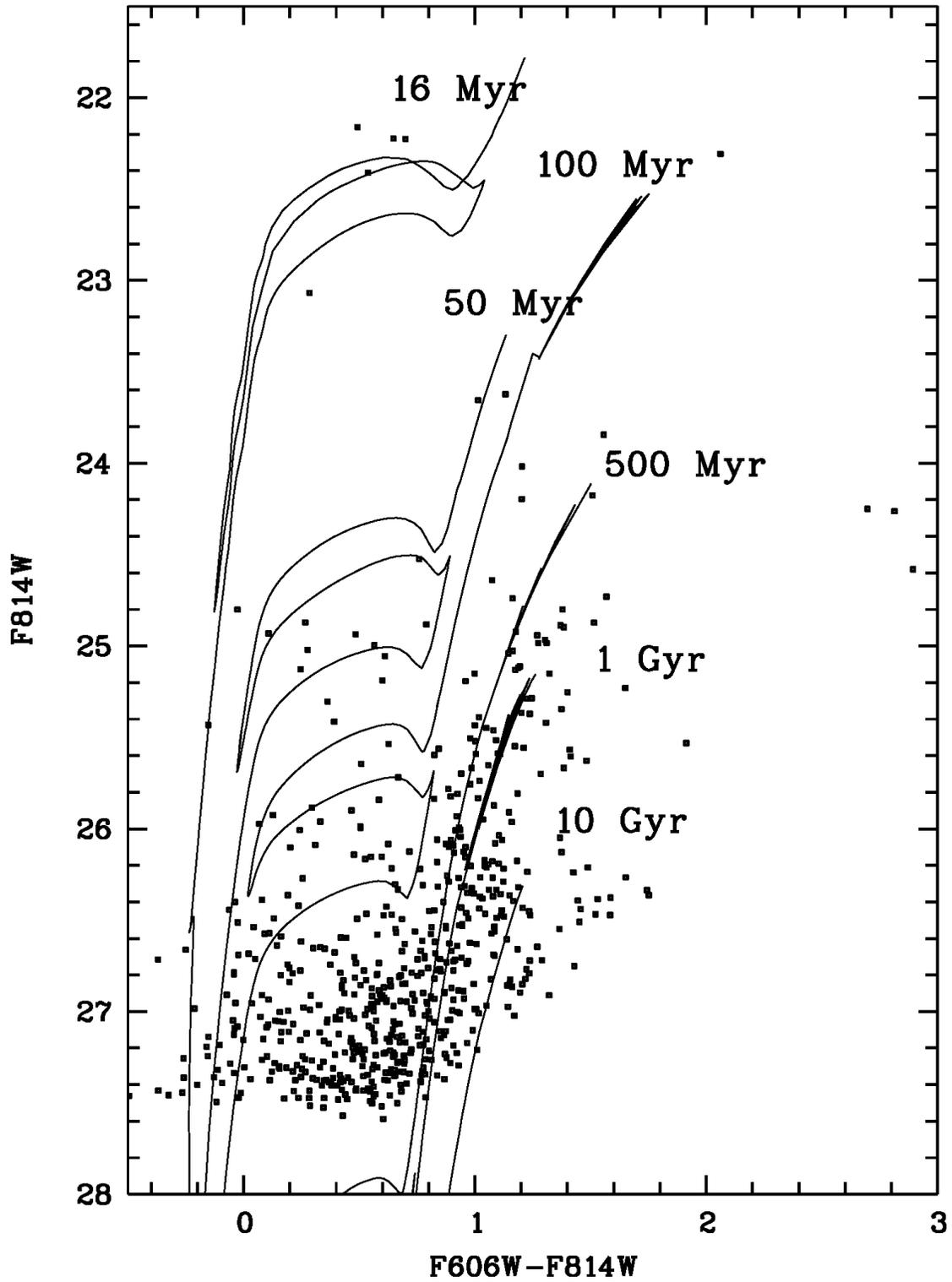}
\caption{Stellar isochrone tracks imposed on the observed color magnitude diagram of Coma P.}
\label{isochrones}
\end{figure*}

Isochrones that stars would lie along at fixed age \citep{2012MNRAS.427..127B} are plotted on the color-magnitude diagram analyzed by LM and DM in Figure~\ref{isochrones}. The isochrone at 16~Myr is speculative; it is motivated by only a few stars that could lie in the foreground. The $50-100$~Myr isochrones are better grounded and would source the UV flux detected by GALEX. Most interesting, though, are the $0.5-1$~Gyr isochrones that describe the AGB population.

These fits assume metallicity $Z=0.001$.  With this choice, the RGB is bracketed by ages of 1 and 10 Gyr.  Given the sparseness of the representation on the RGB, little can be said of metallicity beyond that it is substantially sub-solar.

The ratio of AGB stars to RGB stars is abnormally elevated.  In our experience, our only comparable is the dwarf d0959+68 close to M81 \citep{2013AJ....146..126C}.  This object is suspected to be a ``tidal" dwarf that has formed out of debris related to the interaction between M81, M82, and NGC~3077.  The overwhelming stellar population is suspected to be young.  We do not suggest that the situation with Coma~P is so extreme.  Nevertheless, an elevated ratio of AGB to RGB stars is an indication of substantial star formation within the last few Gyr \citep{2011AJ....141..106J}.  The prominence of the AGB in Coma~P is so pronounced that an intermediate age population of 0.5 to a few Gyr may be dominant.  An ancient population remains to be identified. 

\section{Discussion}

\citet{2018AJ....155...65B} describe the properties of the HI distribution in considerable detail.  The HI velocity field is disjoint, as could be explained by the collision between two units.  There are enhanced stellar densities near the centroids of the two components seen in the HST images and implied by the far and near ultraviolet light in GALEX images.  Our revision of the distance to Coma~P does not particularly affect the scenario of Ball et al. except in the implication that the system is twice as large.  It would lie at the top end in terms of both HI mass and HI diameter within the sample discussed by \citet{2018AJ....155...65B} in connection with their Fig.~16.  Perhaps of relevance to the collision idea is the proximity of HI clouds with no revealed optical counterparts.  AGC 229384 with a velocity differential of $-34$~\kms\ lies 30 kpc away given our distance and AGC 229383, itself shredded into two parts and with a velocity differential of $-61$~\kms, lies 50 kpc away.

We review properties of Coma~P that need to be reconciled.  There is no evidence of ongoing star formation but the UV flux and brightest stars on the main sequence suggest there was activity as recent as 50 Myr ago.  The well populated AGB indicates that there was substantial star formation $0.5-1$~Gyr ago.  The relatively impoverished RGB is consistent with the interesting proposition that any earlier star formation was limited.  The blue color of the TRGB combined with implications of relatively young RGB ages suggests a low metallicity in Coma~P.  Plausibly, an episode of star formation was triggered by the merger of two gas clouds with few or no stars, a possibility supported by the morphology and kinematics of the observed neutral hydrogen. 

Our revision of the distance to Coma~P eliminates the concern regarding its velocity.  Even at 10.9~Mpc, the velocity is a substantial excursion from cosmic expansion but at this distance the system is within the clutches of the Virgo Cluster and its deviant motion is comfortably explained by the detailed flow model of \citet{2017ApJ...850..207S}.  The uncertainty in the distance is large ($9\%$) compared to most TRGB determinations ($5\%$) because of the sparseness of the RGB representation.  Deeper observations could be made but at considerable expense.  It would be more fruitful to obtain distances to either or both of the near neighbors in projection and velocity, NGC~4561 and IC~3605.  We predict that they, too, will be at distances of $\sim11$~Mpc. 

\bigskip\noindent
This research is supported by the Space Telescope Science Institute grant HST-AR-14319.  IK, DM and LM acknowledge the support of the Russian Science Foundation grant 14--12--00965.

\bibliography{paper}
\bibliographystyle{aasjournal}

\end{document}